\documentclass[floatfix,showpacs,amsmath,amssymb,letterpaper,groupaddresses,superscriptaddress]{article}
\usepackage{graphicx}
\usepackage{latexsym}
\usepackage{amsmath}
\usepackage{amssymb}
\usepackage{epsfig}
\usepackage{graphicx}
\usepackage{graphicx}
\usepackage{amssymb,amsmath,times}
\usepackage{hyperref}
\usepackage{color}

\def\a{\alpha}
\def\b{\beta}

\def\be{\begin{equation}}
\def\ee{\end{equation}}
\def\ba{\begin{eqnarray}}
\def\ea{\end{eqnarray}}
\def\la{\langle}
\def\ra{\rangle}

\def\hh{\hskip 2cm}
\def\lo{\longrightarrow}

\def\k{\hat{{\bf k}}}
\def\m{\hat{{\bf m}}}
\def\n{\hat{{\bf n}}}
\def\x{\hat{{\bf x}}}
\def\z{\hat{{\bf z}}}
\def\y{\hat{{\bf y}}}

\usepackage{epsfig}
\usepackage{graphicx}
\usepackage{amssymb,amsmath,times,cite}

\begin{document}

\begin{center}
{\Large \bf Cohering and  de-cohering  power of quantum channels}\\

\vspace{1cm} \hspace{5mm}
Azam Mani \ \ and\ \  Vahid Karimipour
\vspace{5mm}

Department of Physics,\\ Sharif University of Technology,\\
P.O. Box 11155-9161,\\
Tehran, Iran.\\

\end{center}
\vskip 3cm

\begin{abstract}
We introduce the concepts of cohering and de-cohering power of quantum channels. Using the axiomatic defintion of coherence measure, we show that the optimization required for calculations of these measures can be  restricted to pure input states and hence greatly simplified. We then use two examples of this measure, one based on the skew information and the other based on $l_1$ norm, we find the cohering and de-cohering measures of a number of one, two and n-qubit channels. Contrary to a view at first sight, it is seen that quantum channels can have cohering power. It is also shown that a specific property of a qubit unitary map, is that it has equal cohering and de-cohering power in any basis. Finally we derive simple relations between cohering and de-cohering powers of unitary qubit gates and their tensor products, results which have physically interesting implications. 
\end{abstract}


\section{Introduction}

Coherence is a fundamental concept in quantum physics which is closely connected to quantum superposition and quantum correlations. While quantum optics has been the first frame work for understanding the notion of quantum coherence \cite{1Glauber, 2Sudarshan}, this concept is now spread over many fields of science, e.g. from superconductivity \cite{superconduct} to excitation transport in photosynthetic complexes \cite{Photo1, Photo2}.\\

Like entanglement, the quantum coherence is also destroyed during many natural quantum evolutions \cite{MeasureDecoh, 18decoherence2}. It is usually said that a general quantum process, which tends to mix pure input states, de-coheres input state and degrades the original coherence. There are many observations which support this claim, i.e.  the output of a measurement process is always a mixture of projected states each belonging to one result of the measurement. In this way it seems that any measurement completely destroys the coherence of a state \cite{MeasureDecoh}. Generally the output of a quantum channel is an incoherent mixture of states each corresponding to a different error operator acting on the input state.\\

In the absence of precise measures of coherence, these observations only lead to  qualitative statements and claims. If we have definite measures of coherence, then the situation will be different. In that case we will be able to turn the above statements into precise mathematical and physical statements and can even compare the cohering and de-cohering  power of different channels.\\

Motivated by the great success and insight brought about by measures of entanglement defined in quantum information theory, there have been attempts to also define measures of coherence. Some of these measures use the Fisher information \cite{Fisherinfo, Fisher}, some are based on thermodynamic approaches \cite{34ThermoCoherence} and some lead to introducing the concept of catalytic coherence \cite{38Catalytic}. The most recent attempts for quantifying coherence have a resource-based point of view,  i.e. these attempts are based on the fact that coherence is also a resource in quantum information and like any other resource, it can be or should be quantified \cite{Plenio, Girolami, intrinsicRandomness, FidelityTrace, CoherenceWithEnt, GeometricLB}.
 Needless to say, coherence measure, in whatever way it is defined, will depend on the chosen basis, i.e.  a qubit state like $|+\ra:=\frac{1}{\sqrt{2}}(|0\ra+|1\ra)$ is thought to have maximum coherence in the $z-$ basis (the basis $\{|0\ra,|1\ra\}$). Yet the same state may have zero coherene in the $x-$ basis ($\{|+\ra,|-\ra\}$) or some intermediate value in another one.  Nevertheless when we fix a basis, e.g. on some practical ground, then it is meaningful to ask how much a state like $\cos\theta |0\ra + \sin\theta |1\ra$ is coherent. The final result is that one can define $K-$ coherence measures, with respect to a specific basis or observable $K$,  and these measures satisfy reasonable properties. These properties can even be formulated as axioms which any measure of coherence should satisfy \cite{Plenio} (See next section for definitions). Once this is done, then it is quite natural to ask about cohering or de-cohering power of quantum channels and compare them with each other. This study will certainly add another facet to the already rich and interesting subject of quantum channels or completely positive maps. \\ 

What we will do in this paper is the following: In section (\ref{review}) we review the basic axioms for a coherence measure \cite{Plenio}. In section (\ref{twomeasures}) we  review two basic concrete measures which will be used in the sequel,  namely measure based on $l_1$ norm of off-diagonal elements \cite{Plenio} and one based on the skew information \cite{Girolami}. The two measures of coherence, although qualitatively equivalent, are slightly different quantatively. When it comes to concrete calculations, each of these measures have its own drawback as we will see.  We use one or the other depending on the ease with which they can lead to closed forms and also for illustrating their equivalance.\\

In section (\ref{definitions}) we define our  measaures for  cohering and de-cohering power of an arbitrary quantum channel. The cohereing power is naturally defined by finding the maximum amount of coherence a quantum channel can produce when a completely incoherent state is given to it as an input. On the other hand, the de-cohering power of a quantum channel is defined to be the amount by which the coherence of a maximally coherent state is reduced when it passes through the channel. \\
      
These definitions require optimizaitons on the input and output state of the given quantum channel. We will use the axiomatic properties of the coherence measure, which should apply to any specific measure of coherence, and show that the optimization over the input state can be restricted to pure states, hence greatly simplifying the optimization problems. 
In section (\ref{qubitexamples}) we use these two measures to calculate the cohering and de-cohering power of a number of qubit channels. The results are shown in figures  (\ref{fig:fig1}), and (\ref{fig:CoheringPowerBitFlip}). Intersetingly we show that it is a distinctive property of untiary maps that their cohering and de-cohering powers are equal in any basis. Finally in section (\ref{higher}), we prove several general theorems on a class of higher dimensional channels which are of particular interest.  The first is a simple equality which relates the cohering power of a unitary gate $u$ to that of its tensor product $u^{\otimes}$ and the second is a simple inequality relating the decohering power of the two gates. Corollaries of these two theorems relate the powers of individual gates to their tensor products.  \\

We hope to have laid down a path for comparing the cohering and decohering powers of quantum maps and quantum channels, a new window which has been opened in light of quantitative definitions of coherence.  When quantum states are evolved in various complex processes, i.e. in photosynthesis, we can now monitor their coherence according to their passing through different quantum channels. 

\section{Coherence measures of states}\label{review}

The resource based view for quantifying quantum coherence was first introduced in \cite{Plenio}, where the authors have presented some well desired properties for the coherence measures.
The first step for defining a coherence measure is to agree which states of the $d-$ dimensional Hilbert space $\mathcal{H}$ are incoherent. The natural method is to fix a specific basis $\{|i\rangle \ i=1,...,d\}$ of the Hilbert space, and define the set of incoherent states $\mathcal{I}$ to be all the density matrices that are diagonal in this basis:
\be
\rho_{inc}=\sum_i p_i |i\rangle\langle i|.
\ee
The next step is to define the incoherent operations, the operations that do not create coherence when acting on  incoherent states. So the quantum operation $\rho \rightarrow \sum_n \hat{\mathcal{K}}_n \rho \hat{{\mathcal{K}}}_n^\dagger$ (where $\sum_n \hat{{\mathcal{K}}}_n^\dagger  \hat{\mathcal{K}}_n =I$) is an incoherent map if the condition $\hat{\mathcal{K}}_n \mathcal{I} \hat{{\mathcal{K}}}_n^\dagger \subset \mathcal{I}$ is satisfied for all $n$. Equipped with these definitions, any proper coherence measure $C$ is required to satisfy the following conditions \cite{Plenio}:
\begin{itemize}
\item[(C1)] $C(\rho)=0$ iff $\rho \in \mathcal{I}$.
\item[(C2)] Monotonicity under incoherent selective measurements on average: $C(\rho) \geq \sum_n p_n C(\rho_n)$, where $\rho_n ={ \hat{\mathcal{K}}_n \rho \hat{{\mathcal{K}}}_n^\dagger } / {p_n}$ and $p_n= tr \left( \hat{\mathcal{K}}_n \rho \hat{{\mathcal{K}}}_n^\dagger \right) $, with $\sum_n \hat{{\mathcal{K}}}_n^\dagger  \hat{\mathcal{K}}_n =I$ and $\hat{\mathcal{K}}_n \mathcal{I} \hat{{\mathcal{K}}}_n^\dagger \subset \mathcal{I}$.
\item[(C3)] Non-increasing under mixing of states (convexity): $C(\sum_n p_n \rho_n ) \leq \sum_n p_n C(\rho_n)$, for any set of states $\{ \rho_n \}$ and probability distribution $\{ p_n \}$.
\end{itemize}

Thereafter, in view of these properties several coherence measures have been defined \cite{Plenio, Girolami, intrinsicRandomness, FidelityTrace, CoherenceWithEnt, GeometricLB}. Some of these measures use the total $l_1$ norm of off-diagonal elements of the state to quantify the coherence \cite{Plenio}, some belong to the set of distance measures \cite{FidelityTrace}, some use quantum entanglement for quantifying coherence \cite{CoherenceWithEnt}, and finally some other works try to find a lower bound for quantum coherence \cite{GeometricLB}. In this work we use two of these measures, namely the $l_1$ norm of off-diagonal elements \cite{Plenio} and the one based on skew information, recently introduced in \cite{Girolami}.

\subsection{Two specific coherence measures of states}\label{twomeasures}
We will proceed with two measures of coherence as follows, both of which satisfy the properties (C1) to (C3). Both are applicable to states of arbitrary dimensions, however for qubits, they find particularly simple forms.\\

{\bf Remarks:}\\

1- We use a letter $K$ to denote both an observable and the basis of its eignvectors, $\{|k\ra\}$. When using the $l_1$ norm, it is the latter which is implied. \\

2- In the sequel, when we use words like coherence, or cohering power and the like, we always mean $K-$coherence or $K-$cohering power, for the sake of simplicity we do not write the latter explicitly, it is always implied.\\

{\bf \textit{Definition 1: $l_1$ norm of off diagonal elements as a measure of coherence \cite{Plenio}:}}\\

Consider the state $\rho$ and a basis $K:=\{|i\ra\}$. The coherence of this state with respect to this measure is defined as
\be\label{lone}
{C^1}_{K}(\rho):=\sum_{i\ne j}|\la i|\rho|j\ra|.
\ee 
It satisfies all the properties (C1) to (C3). We use the notation ${C^1}_K$ to stress its dependenc on the basis $K$ and to differentiate it from $C_K$ which is defined next. It can be written in the form

\be\label{ltwo}
{C^1}_{K}(\rho):=\sum_{i, j}|\la i|\rho|j\ra|-1,
\ee 
where we have used positivity of $\la i|\rho|i\ra$ and the fact that $tr(\rho)=1$.  In general for a $d-$ dimensional system it satisfies 
\be\label{lthree}
0\leq C^1_{K}(\rho) \leq d-1,
\ee
the maximum value being achieved, for a uniform superposition of basis states of $K$, i.e. $|\psi\ra=\frac{1}{\sqrt{d}}\sum_{i=1}^d |i\ra$. \\

For a qubit state $\rho=\frac{1}{2}(I+{\bf r}\cdot \sigma)$, it is readily found that 
\be\label{c1coherence}
{C^1}_{K}\left(\frac{1}{2}(I+{\bf r}\cdot \sigma)\right)=r\sqrt{1-(\hat{{\bf r}}\cdot \hat{{\bf k}})^2},
\ee 
where $\hat{\bf r}$ is the unit vector $\frac{{\bf r}}{r}$.\\

{\bf \textit{Definition 2: Skew information as a  measure of coherence \cite{Girolami}:}}\\

Let $K$ be an observable with spectrum $\{|k_i \rangle \}$. Then a measure of coherence of a state $\rho$ with respect to this observable, is defined as 
\be \label{GiroCohMeasure}
C_K (\rho) := I \left( \rho , K \right) = - \frac{1}{2}\  tr\left[ \left[ \sqrt{\rho} , K \right]^ 2 \right].
\ee
This coherence measure  introduces a framework for measuring quantum coherence in finite dimensional systems. Moreover it has the extra desirable property that, for any finite-dimensional system, it can be determined by two programmable measurements on an ancillary qubit \cite{Girolami}.If $\rho=|\psi\ra\la \psi|$ is a pure state, then this measure simply reduces to the variance of the observable $K$ for that state, 
\be
C_K(|\psi\ra\la \psi|)=\la \psi|K^2|\psi\ra-\la \psi|K|\psi\ra^2.
\ee
It should be noted that the observable measure of coherence depends not only on the eigenbasis of the operator $K$, but also on its eigenvalues. For the qubit case however, it depends only on the basis and not on the eigenvalues. In fact, the most general qubit observable has the form of $K=\alpha I + \beta \ \vec{\sigma}.\k$, where $I$ is the identity operator, $\vec{\sigma}= (\sigma_x , \sigma_y , \sigma_z)$ denotes the vector of Pauli matrices and $\k$ is a unit vector. From the definition (\ref{GiroCohMeasure}), it is seen that 
\be
C_K(\rho)= \beta ^2 C_{\vec{\sigma}.\k}(\rho),
\ee
which means that  for qubit states once the direction of the measurement $\k$ is fixed, the coherence measures for all observables are proportional to each other.  Hence, for qubit states, it is sufficient to calculate the coherence $C_{\vec{\sigma}.\k}$. For simplicity, in the reminder of this section we use the notation $C_{\k}$ instead of $C_{\vec{\sigma}.\k}$, and call it the $\k-$ coherence. Thus for a qubit state $\rho$, the $\k-$coherence is defined as:

\be
C_{\k}(\rho):=-\frac{1}{2}tr\left[ \left[ \sqrt{\rho} , \vec{\sigma} \cdot {\k} \right]^ 2 \right].
\ee

For the particular case of qubit states this leads to  a closed formula 
\be \label{qubitcoherence}
C_{\hat{k}}\left(\frac{1}{2}(I+{\bf r}\cdot \sigma)\right)=\left( 1-\sqrt{1-{ r}^2} \right) \left( 1- (\bf\hat{\bf r} \cdot \k)^2 \right),
\ee
where $r^2=\bf{r} \cdot \bf{r}$ and ${\bf{\hat{r}}}=\frac{{\bf{r}}}{r}$. \\

For pure qubit states, (when $r=1$), the two measures are very simply related, that is $C_K(|\psi\ra)={C^1}_K(|\psi\ra)^2$. 
Both formulas (\ref{c1coherence}) and (\ref{qubitcoherence}) nicely show how the degree of mixedness of the state (the first factor) ($r $ or $1-\sqrt{1-r^2}$)  and the value of off-diagonal elements of the density matrix (the second factor) play a role in the $\k-$ coherence.  In fact it shows that once the direction $\k$ is fixed, the most coherent states lie on the equatorial plane perpendicular to that direction and the more pure are these states the more coherent they are. 
More concretely,
all states of the form $|\psi\rangle=\frac{1}{\sqrt{2}}\left( |\k+\rangle + e^{i \Omega} |\k-\rangle \right)$ are maximally $\k-$coherent, where $|\k\pm\ra$ are the eigenstates of $\vec{\sigma} \cdot \k$.  It is also seen that for qubit states, 
\be\label{bound}
\forall \  \k \ ,  \ \ \ \  0\leq C_{\k}(\rho)\leq 1,\ \ \ \ \  for \ \ \ \ d=2.
\ee

{\bf Remark:} \textit{Note that the for higher dimensional states, the measure based on the $l_1$ norm depends on the basis that we choose, while the measure based on skew information depends on the observable $K$ (its norm or eigenvalues). In the sequel we use $K$ to refer to a basis when we use the $l_1$ norm and to an observable when we use the skew information as a measure. In general, the two measures can be compared meaningfully only when the latter is normalized in a suitable way. }\\

We are now in a position to use these two measures to calculate the cohering and de-cohering power a number of quantum channels.   \\

We use one or the other depending on the ease with which they can lead to closed forms and also for illustrating their equivalance. In some cases, namely the unitary channel and specifically the Hadamard gate, we use both measures. When it comes to concrete calculations, each of these measures have its own difficulty. For example the measure based on skew information requires taking the square root of a density matrix, which unless $\rho$ is pure is difficult for high dimensional systems, on the other hand, the measure based on the $l_1$ norm leads to cumbersome calculations even for pure states of high dimensionsl due to the absolute values in the sum (\ref{lone}).  
For general qubit states, it is seem from (\ref{c1coherence}) and (\ref{qubitcoherence}), that the two measure of coherence lead to qualitatively similar results. The main difference between the two measures is that the degree of mixedness of a state is measured by $r$ in (\ref{c1coherence}) and almost identically by $1-\sqrt{1-r^2}$ in (\ref{qubitcoherence}). In the next sections we calculate the cohering and de-cohering power of a number of quantum channels. \\

\section{Cohering and de-cohering power of  quantum channels}\label{definitions}
Having a suitable measure to quantify the coherence of quantum states, a natural question is what is the power of a quantum channel for creating or destroying coherence of input quantum states. In this section the definitions are proposed for general channels. The following defintions and the subsequent considerations in this section are valid for any type of coherence measure that one may use for states. \\

{\bf \textit{Definition 3: Cohering power of a channel:}}\\

 For a quantum channel $\mathcal{E}$, we define the cohering power as:
\be \label{coheringDef}
\mathcal{C}_K(\mathcal{E}) := \max_{\rho \in \mathcal{I}} \{ C_K(\mathcal{E}(\rho)) - C_K(\rho) \} = \max_{\rho \in \mathcal{I}} C_K(\mathcal{E}(\rho)) ,
\ee
in which $C_K$ denotes any coherence measure, $\mathcal{I}$ is the set of incoherent states and in the second equality we have used the fact that for an incoherent state $C_K(\rho)=0$.  The definition (\ref{coheringDef}) implies that the cohering power of a channel is the maximum amount of coherence that it creates when acting on
 a completely incoherent state. In a similar way, we can define the decohering power as:\\

{\bf \textit{Definition 4: Decohering power of a channel:}}\\

For a quantum channel $\mathcal{E}$, the decohering power is defined as:
\be \label{decoheringDef}
\mathcal{D}_K(\mathcal{E}) := \max_{\rho \in \mathcal{M}} \{ C_K(\rho) - C_K(\mathcal{E}(\rho)) \},
\ee
where again $C_K$ stands for any coherence measure  and $\mathcal{M}$ is the set of maximally coherent states.
According to this definition, the decohering power of the channel $\mathcal{E}$ is the maximum amount by which it reduces the coherence of a maximally coherent state.\\

Using the properties of the coherence measure $C$ one can easily show that the optimizations in equations (\ref{coheringDef}) and (\ref{decoheringDef}) can be reduced to simple maximizations over a small set of parameters.  To this end we note that any $K-$incoherent state is diagonal in the eigenbasis of $K$, i.e. $\rho_{inc} \in \mathcal{I}$ iff $\rho_{inc}=\sum_i p_i |k_i\rangle \langle k_i |$. Therefore using the convexity property of the coherence measures (C3), we find
\ba \label{CohPowerProof}
\mathcal{C}_{K} \left( \mathcal{E} (\rho_{inc}) \right) &=& C_K \left( \mathcal{E} (\sum_i p_i |k_i\rangle \langle k_i |) \right) \cr
&=& C_K \left( \sum_i p_i \mathcal{E} ( |k_i\rangle \langle k_i |) \right) \cr
&\leq & \sum_i p_i C_K \left( \mathcal{E}(|k_i\rangle \langle k_i |) \right) \cr
&\leq & C_K \left(\mathcal{E}(|k^\star \rangle\langle k^\star |) \right),
\ea
where $|k^\star\ra$ is the basis vector which has the largest contribution on the right hand side. 
Since $|k^\star\rangle$ is itself a $K-$incoherent state, the upper bound (\ref{CohPowerProof}) is achieved and hence 
\be\label{CohPower}
\mathcal{C}_{K} \left( \mathcal{E} \right)
= C_K \left(\mathcal{E}(|k^\star \rangle\langle k^\star |) \right)
= \max_i C_K \left( \mathcal{E}(|k_i\rangle \langle k_i |) \right),
\ee
i. e. the continuous optimization in (\ref{coheringDef}) reduces to a simple discrete maximization.\\

We can also simplify equation (\ref{decoheringDef}). We note from (C3)  that a coherence measure should not  increase under mixing and hence all maximally coherent states are pure ones. This means that in equation (\ref{decoheringDef}), we can take the input state to be a pure state which has maximum $K-$ coherence, i.e.
\be\label{DeCohPower}
\mathcal{D}_{K} \left( \mathcal{E} \right) = C_K(|\psi\ra\la \psi|)- \min_{|\psi\rangle \in \mathcal{M}} C_K \left(\mathcal{E}(|\psi \rangle\langle \psi |) \right),
\ee
where minimization is performed over all maximally coherent states in the basis of $\hat{K}$. \\

We should stress that the above considerations are valid for any type of coherence measure and any quantum channels in any dimension. They are not specific to qubit channels.  Finally we should note that in view of the our definitions the following bounds are valid:

\be\label{bound2}
0\leq \mathcal{C}_{\bf k}(\mathcal{E})\leq 1, \ \ \ \ \  0\leq \mathcal{D}_{\bf k}(\mathcal{E})\leq 1, \ \ \ \  {\rm\ for } \ \ \ d=2, 
\ee
and

\be\label{boundONE}
0\leq {\mathcal{C}^1}_ K(\mathcal{E})\leq d-1, \ \ \ \ \  0\leq \mathcal{D}_K(\mathcal{E})\leq d-1, \ \ \ \  {\rm\ for\ \ \  any }\ \  \ d. 
\ee

After these general definitions, we are now ready to apply them to specific channels.

\section{Examples of cohering and de-cohering power of channels}\label{examples}

Having defined the cohering and decohering power of quantum channels, we go on to calculate these powers for some specific channels and we will see that these measures behave in a physically expected manner. We start from the simplest example, namely the unitary operations and then go on to qubit depolarizing channel and bit-flip channel. We also consider gates of the form $u^{\otimes n}$ and the CNOT gate as illustrative examples of higher dimensional gates. \\

\section{Qubit channels}\label{qubitexamples}
Using the closed formula (\ref{qubitcoherence}), we can find closed expressions for cohering and de-cohering power of qubit channels. Let $\rho=|m\ra\la m|=\frac{1}{2}(I+\hat{{\bf m}}\cdot \sigma)$ be the pure input state of a channel $\mathcal{E}$. The input state is thus represented by a Bloch vector of unit length $\m$. We then have $\mathcal{E}(\rho)=\frac{1}{2}(I+{\bf m'}\cdot \sigma)$, where ${\bf m'}$ is not necessarily a unit vector. Therefore
\be\label{qubitFinalcoherence}
C_{\hat{{\bf k}}}(\mathcal{E}(\rho))=\left( 1-\sqrt{1-{ m'}^2} \right) \left( 1- (\hat{{\bf m'}} . \k)^2 \right).
\ee
This quantity depends on the input Bloch vector $\m$, the coherence direction ${\k}$ and the parameters of the channel $\mathcal{E}$. Let us denote it by $F_{\mathcal{E}}(\hat{{\bf m}}, \hat{{\bf k}})$:

\be \label{Fdefinition}
F_{\mathcal{E}}(\hat{{\bf m}}, \hat{{\bf k}})=\left( 1-\sqrt{1-{ m'}^2} \right) \left( 1- (\hat{{\bf m'}} . \k)^2 \right),\ee
where ${\bf m'}$ is the output Bloch vector of the channel $\mathcal{E}$ for the input Bloch unit vector $\hat{{\bf m}}$ and $\hat{\bf m'}$ is its unit vector.\\

For calculating the cohering power, this quantity should be maximized over input pure states which are incoherent in the  ${\k}-$ basis, i.e. ${\m}=\pm \k$. Hence only a two-fold maximization is necessary and one simply arrives at: 
\be\label{qubitCohPow}
\mathcal{C}_{\hat{{\bf k}}}(\mathcal{E})= \max \left\{ F_{\mathcal{E}}(\hat{{\bf k}},\hat{{\bf k}}), F_{\mathcal{E}}(-\hat{{\bf k}},\hat{{\bf k}}) \right\}.
\ee

For calculating the de-cohering power, this quantity should be minimized over input pure states which are the most coherent in the  $\hat{\bf k}-$ basis, i.e. $\hat{\bf m}\perp \hat{\bf k}$. That is, the input state is an equatorial pure state with respect to the direction $\hat{{\bf k}}$, in other words, the input state is of the form $|\psi\ra=\frac{1}{\sqrt{2}}(|\k+\ra+e^{i\Omega}|\k-\ra)$, where $|\k_\pm\ra$ are the two eigenvectors of $\vec{\sigma}\cdot \hat{{\bf k}}$.  In this case the minimization (\ref{DeCohPower}) is effectively performed on the relative phase $\Omega$. 
Thus we have 

\be \label{qubitDeCohPow}
\mathcal{D}_{\hat{k}}(\mathcal{E})=1-\min_{{\m}, {\m}\cdot {\k}=0} F({\m},{\k}).
\ee
We now turn to specific examples of qubit channels.

\subsection{A unitary channel}
\begin{figure}
\begin{center}
\vskip -3 mm
\includegraphics[width=9cm, height=7.5cm]{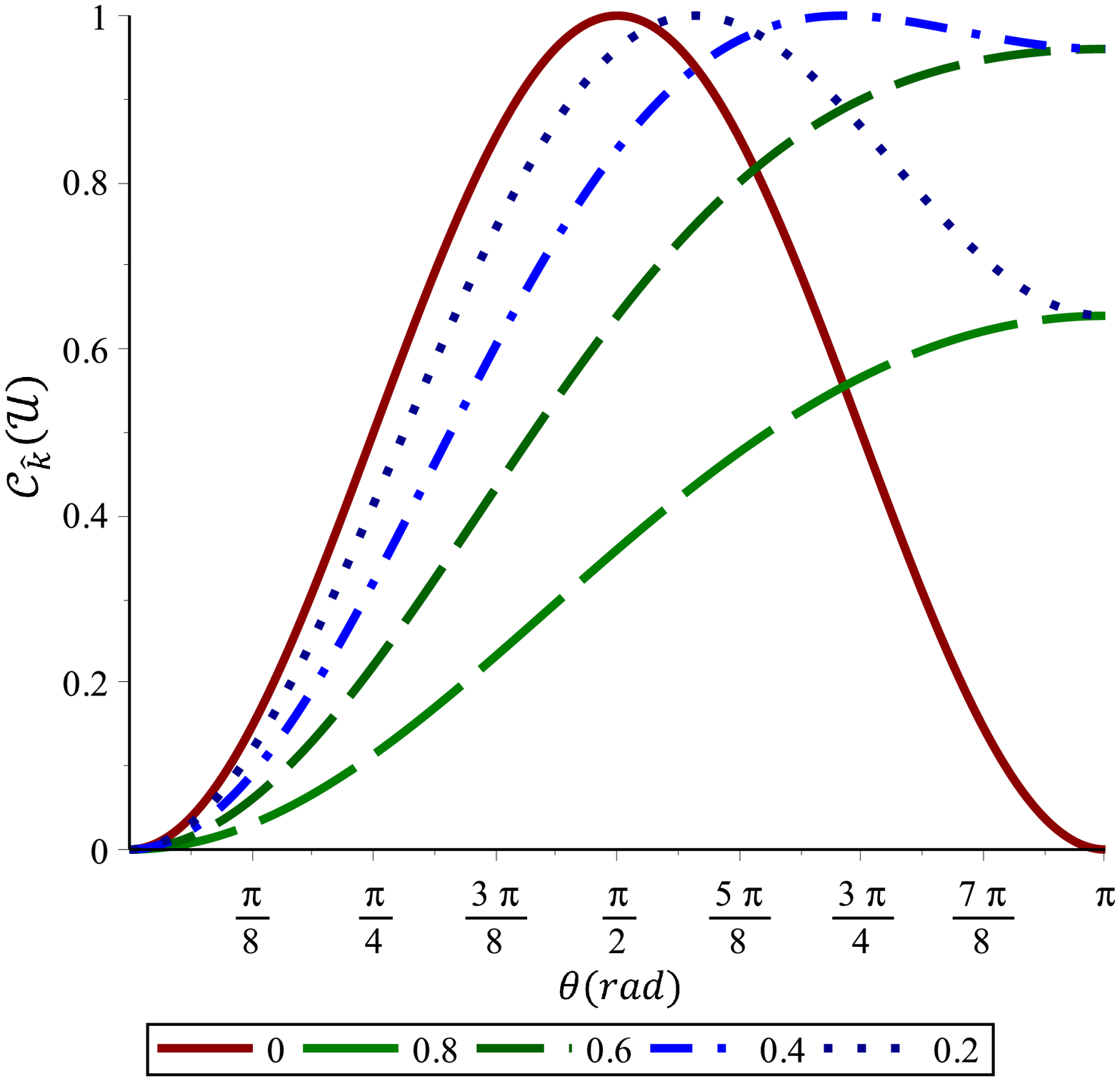}
\caption{(Color online) $\k-$ Cohering and decohering power of a unitary channel  $U=e^{i\frac{\theta}{2} \n\cdot \sigma}$ ( a $ \theta $ rotation around $\n$), as a function of the angle of rotation, for various values of $\k\cdot \n$. The numbers denote the value of $\k\cdot \hat{\bf n}$. Without loss of generality, we can take $\n$ to be along the $z$ direction. Then the numbers would correspond to different values of $\k\cdot \hat{\bf z}$. The cohering power is a dimensionless quantity.}
\label{fig:fig1}
\end{center}
\end{figure}

Depending on the basis we choose for measuring coherence, unitary channels can create coherence or destroy it. For example the $z-$coherence of the Hadamard gate is maximum since it takes an incoherent pure state $|0\ra$ and turns it into a maximally coherent state $|+\ra:=\frac{1}{\sqrt{2}}(|0\ra+|1\ra)$. Conversely it turns $|+\ra$ into $|0\ra$, so its de-cohering power is also maximum. We now ask: Does every unitary channel has this property? What is the cohering and de-cohering powers of a general unitary qubit channel $\mathcal{U}(\rho)=U\rho U^{\dagger}$, where $U=e^{i  \frac{\theta}{2}\n\cdot \sigma}$. In view of (\ref{qubitFinalcoherence}), the starting point is to see how the Bloch unit vector changes under this map. Being a unitary map, the input vector rotates around $\n$ by an angle $\theta$, i.e. and its norm is preserved: 

\be
{\bf m}'= \cos \theta \m + \sin\theta (\m\times \n)+(1-\cos \theta) (\m\cdot \n)\n.
\ee
Therefore from (\ref{qubitCohPow}) we find 

\be\label{coU}
\mathcal{C}_{\k}(\mathcal{U})=\left(1-[\cos \theta + (1-\cos \theta)(\k\cdot \n)^2]^2\right).
\ee
Figure (\ref{fig:fig1}), shows the cohering power of the unitary channel $U=e^{i \frac{\theta}{2} \n\cdot \sigma}$ as a function of the angle of rotation in various bases. For such a channel when $\k$ is parallel to $\n$, no coherence is produced but as the angle between $\n$ and $\k$ increases, the cohering power also increases. Maximum cohering power is achieved when $\k$ is perpendicular to the axis of rotation and the angle of rotation is $\frac{\pi}{2}$. 
\\

\begin{figure}
\begin{center}
\vskip -3 mm
\includegraphics[width=8cm, height=7.5cm]{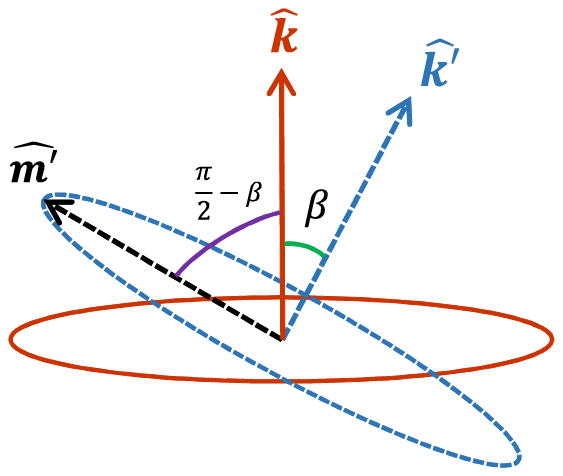}
\caption{(Color online) The unitary operation $U=e^{i\frac{\theta}{2} \n\cdot \sigma}$ rotates the vectors rigidly. The solid red vector shows the basis $\k$ for which we want to calculate the powers and the solid red circle shows the space of unit vectors $\m$, $\m\perp \k$. The dashed blue vector shows $\k'$ which is the image of $\k$ under the action of $U$ and the dashed blue circle represent the vectors $\m'$ which are images of the vectors $\m$. }
\label{fig:unitary}
\end{center}
\end{figure}

In principle and in view of the general definitions (\ref{coheringDef}) and (\ref{decoheringDef}), the cohering and de-cohering powers of unitary channels can be different. However as   we now show, unit unitary channels have the distinctive property that their cohering and de-cohering powers are equal in any basis. To prove this we remind if $\k$ is the Bloch vector of the observable $K$, then the K-cohering power of the unitary channel is given by:

\be\label{CU}
\mathcal{C}_{\k}(\mathcal{U})= 1 - (\k' \cdot \k)^2=\sin^2 \beta,
\ee

where the vectors $\k'$ is the result of rotation of  $\k$ by the unitary map, and $\beta$ is the angle between the two vectors, figure (\ref{fig:unitary}). Using the basic definitions (\ref{decoheringDef}) and (\ref{qubitcoherence}), we note that 
the de-cohering power of this unitary map in the same basis is given by:  
 
\be\label{DU}
\mathcal{D}_{\k}(\mathcal{U})= \max_{\m, \m\cdot\k=0} (\m' \cdot \k)^2,
\ee
where we have used the unitary property which maps pure states to pure states. 
To prove equality of (\ref{CU}) and (\ref{DU}) is now a simple problem of geometry. We use the fact that the unitary channel rotates the vectors rigidly and it does not change the angle between the directions, so we can easily rephrase equation (\ref{DU}) to:
\be\label{DU2}
\mathcal{D}_{\k}(\mathcal{U})= \max_{\m', \m'\cdot\k'=0} (\m' \cdot \k)^2=\cos^2 (\frac{\pi}{2}-\beta)=\sin^2\beta,
\ee
where in the last equality we have used the relations of vectors in figure (\ref{fig:unitary}).  Hence we have shown that for any unitary channel the cohering and de-cohering powers are equal in any basis:

\be\label{CD}
\mathcal{D}_{\k}(\mathcal{U})=\mathcal{C}_{\k}(\mathcal{U}).
\ee 
 This is a distinctive property of the unitary maps for qubits and may not hold for other measures of coherence or in other dimensions. \\

 {\bf Example 1:} Let us illustrate these with a familiar and important example, the Hadamard gate. First we note that the cohering and de-cohering power of a unitary gate does not change if we multiply it by a total phase. So, instead of the Hadamard gate, we can take the Hadamard gate to be $H=\frac{1}{\sqrt{2}}\left(\begin{array}{cc} 1 & 1 \\ -1 & 1\end{array}\right)=\frac{1}{\sqrt{2}}(I+i\sigma_y)=e^{i\frac{\pi}{2}\sigma_y}$. Then according to (\ref{coU}), we have  
\be
C_{\k}(H)=1-{k_2}^2
\ee
where $k_2$ is the second component of $\k$. This means that 
\be\label{ch}
C_{\x}(H)=C_{\z}(H)=1 \ \ \ \ , \ \ \ \ C_{\y}(H)=0. 
\ee
Let us verify these explicitly. For the $\z-$ coherence, the input to the Hadamard gate is $|0\ra$ or $|1\ra$, and the output will be $\frac{1}{\sqrt{2}}(|0\ra\pm|1\ra$, both of which have maximum $\z$-coherence, hence $C_{\z}(H)=1$. Similarly for the $\x-$ coherence, the input to the Hadamard gate is $|+\ra$ or $|-\ra$, and the output will be $|0\ra $ or $|1\ra$ which are of the form $\frac{1}{\sqrt{2}}(|+\ra\pm|-\ra$, both of which have maximum $\x$-coherence, hence $C_{\x}(H)=1$. On the other hand, for the  $\y-$ coherence, the input to the Hadamard gate is $|y_{\pm}\ra=\frac{1}{\sqrt{2}}(|0\ra\pm i|1\ra)$ and the output is found to be be $e^{\pm i \frac{\pi}{4}}|y_{\mp\ra}$, both of which have zero $\y$-coherence, hence $C_{\y}(H)=0$. Note that according to (\ref{CD}), the de-cohering power of Hadamard gate is the same as its cohering power obtained above. \\
Note that if we had normed the $l_1$ norm, in view of its equivalence to the norm, we would have obtained essentially the same results.   

\subsection{Depolarizing channel}
The depolarizing channel is a quantum process which converts any qubit state $\rho$ to:
\be\label{depolarizCH}
{\mathcal{E}}_{dep}(\rho)=(1-p) \rho + p \frac{I}{2}, \hh 0\leq p \leq 1.
\ee

In view of the rotational covariance of this channel, it is obvious that its cohering and de-cohering powers are the same for all directions. It is rather obvious that this channel  has no $z-$cohering power and hence we expect that $$\mathcal{C}_{\bf \hat{k}}(\mathcal{E}_{dep})=0.$$ The reason is quite simple, we feed an incoherent pure state like $|\k_+\ra$ into this channel and at the output we get $(1-p)|\k+\ra\la \k+|+p\frac{I}{2}=(1-\frac{p}{2})|\k+\ra\la \k+|+\frac{p}{2}|\k_-\ra\la \k_-|$, which is again an incoherent state in the ${\k}$ basis. This is reflected in the formula (\ref{qubitCohPow}), since in this case, the output Bloch vector is  ${\bf m}'=(1-p)\hat{{\bf k}}$ and hence $\hat{{\bf m}}'=\hat{{\bf k}}$, leading to a vanishing value for $F_{\mathcal{E}_{dep}}(\pm \hat{{\bf k}},\hat{{\bf k}})$, according to (\ref{Fdefinition}).\\ 

The interesting point is that this channel which cannot create any coherence, nevertheless does not have full de-cohering power. It cannot fully destroy the coherence of maximally coherent states. In fact from (\ref{Fdefinition}) and (\ref{qubitDeCohPow}), and noting that ${\bf m'}=(1-p)\hat{{\bf m}}$ and ${\bf m}'\cdot {\k}=0$ we have 

\be
\mathcal{D}_{\k}(\mathcal{E}_{dep})= \sqrt{1-(1-p)^2}.
\ee
As expected the de-cohering power increase from $0$ for the identity channel to the maximum value of $1$ for a completely depolarizing channel.\\

\subsection{Bit-flip channel}
The Bit-flip channel is the last qubit channel that we study in this paper. Interestingly we will see that this channel has a non-zero cohering power. The channel is defined as

\be
\mathcal{E}_{bf}(\rho)=(1-p) \rho + p \sigma_x \rho \sigma_x, \hh 0 \leq p \leq 1,
\ee
where $\sigma_x$ is the Pauli matrix. It transforms the input Bloch vector $\hat{{\bf m}}$ into

\be\label{m'm}
{\bf m}'=(1-2p)\m +2p (\m \cdot \x) \x. \ee

The first factor of the function $F(\m,\k)$, namely $\sqrt{1-\bf m'\cdot\bf m'}$ is found to be $\sqrt{1-\bf m'\cdot\bf m'}=\sqrt{4p(1-p)(1-(\m\cdot\x)^2)}$. The second factor depends on whether we take $\m=\pm\k$ (for the cohering power) or $\m\cdot\k=0$ (for de-cohering power). We consider these two cases separately.\\

\begin{figure}[tp]
\begin{center}
\vskip -3 mm
\includegraphics[width=9cm, height=7.5cm]{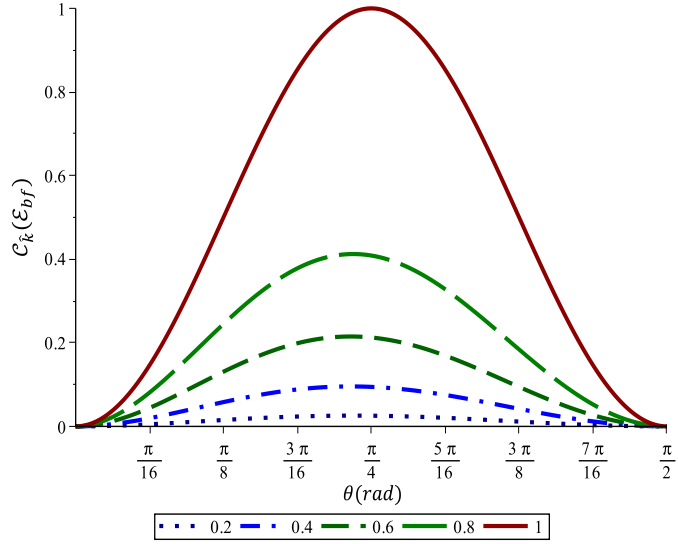}
\caption{(Color online) Cohering power of the bit-flip channel as a function of the direction of coherence for various values of $p$. The parameter $\theta$ is the angle between the $\x-$axis and the $\k-$axis. From bottom to top $p=0.2, \ 0.4, \ 0.6, \ 0.8$ and $p=1$. The cohering power is a dimensionless quantity.} 
\label{fig:CoheringPowerBitFlip}
\end{center}
\end{figure}

\textit{Cohering power:} In this case we take $\m=\pm\k$ and after using (\ref{qubitCohPow}) and some simple algebra, we arrive at

\be
\mathcal{C}_{\k}(\mathcal{E}_{bf})= \left(1-\sqrt{4p(1-p)(1-\eta)}\right)\frac{4p^2\eta(1-\eta)}{1-4p(1-p)(1-\eta)},
\ee
where $\eta:=(\k\cdot \x)^2.$ Note that one can substitute $\x$ with $\hat{{\bf z}}$ to obtain the corresponding result for the phase-flip channel.  Figure (\ref{fig:CoheringPowerBitFlip}) shows the cohering power as a function of $\eta=(\k\cdot\x)^2=:\cos\theta$ for various values of the parameter $p$. Note that $\theta$ is the angle between the $x-$axis and the axis for which we calculate the coherence. Hence for $\theta=0$, figure (\ref{fig:CoheringPowerBitFlip}) shows the $x-$coherence of bit-flip channel for various values of $p$ and for $\theta=\frac{\pi}{2}$, it shows the $z-$coherence of this channel. 
It is seen that $x-$coherence and $z-$coherence of the Bit-Flip channel is zero for all values of $p$. This is exactly what we expect. In fact for the $x-$coherence, a pure incoherent state is untouched by this channel and no coherence is produced. Also for the $z-$coherence, a pure incoherent state like $|0\ra$ is turned into $(1-p)|0\ra\la 0|+p|1\ra\la 1|$ which is again a $\z-$incoherent mixture of states. However the bit-flip channel has non-vanishing  cohering power for other directions. In fact from figure (\ref{fig:CoheringPowerBitFlip}) it is seen that the maximum coherence is produced for $\theta \approx \frac{\pi}{4}$ almost independent of $p$. 
The maximum coherence is achieved for (p=1) which is the unitary channel  $\rho\lo \sigma_x\rho \sigma_x $. Such a channel has the maximum cohering power for $\k = \frac{1}{\sqrt{2}}(\hat{\bf x}+\hat{\bf z})  $ ( $\theta=\frac{\pi}{4}$ in  figure (\ref{fig:CoheringPowerBitFlip})). In fact an incoherent pure state in the $\k$ direction is an eigenstate of $\vec{\sigma} \cdot \k=\frac{1}{\sqrt{2}}\left(\begin{array}{cc} 1 & 1 \\ 1 & -1\end{array}\right)$. The eigenstates of this Hadamard gate are $|\k+\ra:=\cos \frac{\pi}{8}|0\ra + \sin\frac{\pi}{8}|1\ra$ and $ |\k-\ra:=\sin \frac{\pi}{8}|0\ra -\cos\frac{\pi}{8}|1\ra$. When the incoherent state $|\k+\ra$ is fed into the unitary channel $\rho\lo \sigma_{x} \rho \sigma_{x}$, it is converted to 
$\cos \frac{\pi}{8}|1\ra + \sin\frac{\pi}{8}|0\ra=\frac{1}{\sqrt{2}}(|\k+\ra-|\k-\ra)$ which is a maximally $\k-$coherent state. Hence the bit-flip channel for $p=1$ or the unitary operator $|\psi\ra\lo \sigma_x|\psi\ra$ has maximum cohering power in the $\k=\frac{1}{\sqrt{2}}(\hat{\bf x} + \hat{\bf z})$ basis. Obviously the same results hold if one replaces 
$\hat{\bf z}$ with any other vector in the $y-z$ plane. \\

\textit{De-cohering power:} Calculation of the de-cohering power is more complicated than the cohering power since it requires a continoious optimization. 
To this aim we take $\m\cdot \k=0$ in (\ref{Fdefinition}), and by using (\ref{m'm}) and doing simplifications, we find

\be
F(\m,\k)\mid_{\m\cdot\k=0}=:F_{\a,\b}(\xi)=\left(1- \sqrt{\alpha (1-\xi)} \right) \left(1- \frac{\beta \xi}{1-\alpha+ \alpha \xi }\right)
\ee
where  $\alpha= 4 p(1-p)$, $\beta= 4p^2(\k \cdot \x)^2$ and $\xi := (\m \cdot \x)^2$, with $\m$ being all unit vectors that are perpendicular to $\k$. What we have to find is

\be\label{bitDecohPow1}
\mathcal{D}_{\k}(\mathcal{E}_{bf})=1-\min_{\xi} F_{\alpha,\beta}(\xi).
\ee
\\
The calculation is detailed in the appendix and the final result is:
\be \label{bitDecohFinal}
\mathcal{D}_{\hat{k}} \left(\mathcal{E}_{bf}\right)=\left \lbrace
\begin{array}{ccc}
2\sqrt{p(1-p)}  &  \text{\ \ \ \ \ if } (\k \cdot \x)^2 \leq A, \cr
\cr
\frac{4p (\k\cdot\x)^2 (1-p (\k\cdot\x)^2)+\sqrt{4p(1-p)(\k\cdot\x)^2}}{1+\sqrt{4p(1-p)(\k\cdot\x)^2}} & \text{\ \ \ \ \ if } (\k \cdot \x)^2\geq A.
\end{array}\right.
\ee
\\
in which 
\\
\be
A:= \frac{1}{2} \left( \frac{1-p}{p} + \frac{\sqrt{4p(1-p)}}{4p^2} \right)
\ee

We see that the de-coheing power the bit-flip channel depends both on the parameter $p$ of the channel and on the angle between the direction $\x$ and $\k$ for which we calculate the power. Note that this would be expected due to the symmetries of the bit-flip channel. Several special cases of interest are:\\

i) $p\leq \frac{1}{2}$:   In this case the parameter $A$ is always larger than $1$ and hence for any arbitrary direction, the $\hat{k}-$decohering power of the bit-flip channel is always $2\sqrt{p(1-p)}$.\\

ii) $\k =\hat{x}$:  Here, regardless of the value of $A$ (or $p$), the decohering power of the bit-flip channel is $2\sqrt{p(1-p)}$. This  is expected since the Bloch vector of maximally $\hat{x}-$coherent states lies in the $y-z$ plane, and the bit-flip channel only reduces the length of these vectors by a factor $(1-2p)$.\\

iii) $\k=\hat{z}$ (or any vector which is perpendicular to the $\x$ axis): This is the situation that we are interested in the $\z-$decohering power of the bit-flip channel. In this case the input states of the bit-flip channel are maximally $\z-$coherent states and their Bloch vector lie in the $x-y$ plane. The minimum corresponds the the Bloch vector $\y$ which is converted to $(1-2p)\y$ after the action of the bit-flip channel. So as it is also expected due to equation (\ref{bitDecohFinal}), the $\z-$decohering power of the bit-flip channel is $2\sqrt{p(1-p)}$.\\

\section{Higher dimensional channels}\label{higher}
To have an idea of cohering and de-cohering power of higher dimensional channels and leaving out trivial cases like $d$-dimensional depolarizing channel for which cohering power is obviosuly zero,  we investigate two well-known examples which have a vast application in quantum computation information and computation. One is the multi-qubit gate $u^{\otimes n}$, where $u$ is a qubit unitary  and the other is CNOT gate. In the special case where $u=H$, the former is used for creating a uniform superposition of states in the computational basis and the latter is used as an entangling gate. As a byproduct,  this study will let us know the relation of cohering and entangling power. For both of these gates, we will use the definition 3 which is based on $l_1$ norm.

 \subsection{Gates of the form $u^{\otimes n}$}
Consider an $n$ qubit unitary gate $u^{\otimes n}$, we want to find the $K-$ cohering and $K-$ decohering power of this channel, where we use the $l_1$ norm as a measure and  take $K$ to be a product basis of the form $\{|K_r\ra\} = \{|k_{r_1}\cdots k_{r_n}\ra\} $. Denoting the one-qubit basis $\{|k\ra\}$ by $k$, we can then prove the follwing theorem:\\

{\bf Theorem 1}: Using the $l_1$ norm as a measure of coherence, the $K-$cohering power of the gate $u^{\otimes n}$ is related to the $k-$ cohering power of a gate $u$ as follows:

\be\label{cnn}
{C^1}_{{K}}(u^{\otimes n})=\left[{C^1}_{{k}}(u)+1\right]^n-1.
\ee

This is in accord with what we know from the behaviour of Hadamard gate. A single qubit Hadamard gate when acting on a state $|0\ra$, produces a uniform superposition of $|0\ra$ and $|1\ra$, hence it has unit cohering power. In the same way $H^{\otimes n}$ when acting on $|00\cdots 0\ra$  produces a uniform superposition of all computational states, hence it also has a maximum cohering power of $2^n-1$ cohering power which is reflected in the previous formula. \\

{\bf Proof:}
Consider a single qubit gate $u$, and a basis $\{|k_r\ra\}$. Then from (\ref{ltwo}) and (\ref{CohPower}) we have 
\be
{C^1}_k(u) = \max_{k} \sum_{k_i\ne k_j} |\la k_i|u|k\ra\la k|u^\dagger |k_j\ra|=\max_{k} \sum_{k_ i , k_j} |\la k_i|u|k\ra\la k|u^\dagger |k_j\ra|-1.
\ee 

Consider now the coherence power of the gate $u^{\otimes n}$ in the product basis $|K\ra$. The basic steps of the calculations is best understood for the simple case of $n=2$ and can easily be generalized to arbitrary $n$.

\ba\nonumber
{C^1}_{{K}}(u^{\otimes 2}) &=& \max_{k,k'}\sum_{(k_i, k_j)\ne (k_l, k_m)}|\la k_i, k_j|u^{\otimes 2}|k,k'\ra\la k,k'|{u^{\dagger}}^{\otimes 2}|k_l,k_m\ra|\cr
&=& \max_{k,k'}\sum_{(k_i, k_j, k_l,k_m)} |\la k_i, k_j|u^{\otimes 2}|k,k'\ra\la k,k'|{u^{\dagger}}^2|k_l,k_m\ra|-1\cr &=&\left(\max_k\sum_{k_i, k_l}
 |\la k_i|u|k\ra\la k|u^{\dagger}|k_l\ra |\right)\left( \max_{k'}\sum_{k_j, k_m}
 |\la k_j|u|k'\ra\la k'|u^{\dagger}|k_m\ra|\right)-1
\ea
Using (\ref{ltwo}) we can express the right hand side in terms of cohering power of $u$ and arrive at $
C_{{K}}(u^{\otimes 2})=\left[C_{{k}}(u)+1\right]^2-1.
$ 
Straightforward repeation of this calculation leads to (\ref{cnn}) and completes the proof of the theorem.  \\

{\bf Corollary:} With a similar argument, one can directly derive the following relation:

\be
{C^1}_{{K}}\left(\bigotimes_{i=1}^n u_i\right)=\prod_{i=1}^n\left[{C^1}_{{k}}(u_i)+1\right]-1.
\ee

The  decohering power of $u^{\otimes n}$ cannot be simply related to that of the gatge $u$, but we can find useful bounds. \\

{\bf Theorem 2:} Using the $l_1$ norm measure of coherence, the $K-$ decohering power of a gate $u^{\otimes n}$ is related to the $k-$ de-cohering power of a gate $u$ in the following way:\\

\be
{D^1}_K(u^{\otimes n})\geq 2^n-[2-{D^1}_k(u)]^n.
\ee

Note that if ${D^1}(u)=0$, then we also have ${D^1}_K(u^{\otimes n})=0$. So no cohering power is generated by tensor product of gates which have zero cohering power. We also see that a gate which has a unit de-cohering power in a basis $\{|k\ra\}$ (which is the maximum value for a qubit gate) when put in tensor product, will also have a decohering power in the basis $\{|K\ra=|k\ra^{\otimes n}\}$, equal to the largest possible value (which is $2^n-1$ in this case).  \\

{\bf Proof:}
We denote the basis states by $|i\ra,\ |j\ra$ and so on. First we note from (\ref{DeCohPower}) that the de-coherening power of a unitary qubit gate $u$, measured by $l_1$ norm is given by 
\ba
{D^1}_k(u)&=&1-\min_{|\psi\ra\in \mathcal{M}_1} \sum_{i\ne j}|\la i|u|\psi\ra\la \psi|u^\dagger|j\ra|\cr &=& 2-\min_{|\psi\ra\in \mathcal{M}_1}\sum_{i, j}|\la i|u|\psi\ra\la \psi|u^\dagger|k\ra|,
\ea
    where $ \mathcal{M}_1$ is the set of all one qubit maximally coherent pure states.
We now explain the details of the proof for $n=2$, which is readily generalized to arbitrary values of $n$. We use the measures based on $l_1$ norm and  note from definitions (\ref{ltwo}) and (\ref{DeCohPower}) that 

\be
{D^1}_K(u^{\otimes 2})=2^2-1-\left(\min_{|\psi\ra\in \mathcal{M}_2} \sum_{i,j;k,l} \mid\la i,j|u^{\otimes 2}|\psi\ra\la \psi|{u^\dagger}^{\otimes 2}|k,l\ra\mid-1 \right),
\ee
where the minimum is taken over all the two-qubit maximally coherent states $|\psi\ra\in \mathcal{M}_2$ (hence the factor $2^2-1$). We now take a restricted class of maximally coherent states of the form 

\be
|\psi\ra=|\psi_1\ra\otimes|\psi_2\ra \in \mathcal{M}_1\otimes  \mathcal{M}_1
\ee
i.e. where each of the states $|\psi_i\ra$ is a maximally coherent one qubit state, i.e. $|\psi_i\ra=\frac{1}{\sqrt{2}}(|0\ra+e^{i\a_i}|1\ra)$. Clearly this leaves out the maximally coherent entangled states of the form 
$|\psi\ra=\frac{1}{2}(|00\ra+e^{i\a}|01\ra+e^{i\beta}|10\ra+e^{i\gamma}|11\ra)$. Since the coherent states we are considering, is a subclass of all coherent states, we find that 

\ba\nonumber
{D^1}_K(u^{\otimes 2})&\geq &
2^2-\left(\min_{|\psi_1\ra\in \mathcal{M}_1} \sum_{i,k} \mid\la i|u|\psi_1\ra\la \psi_1|{u^\dagger}|k\ra\mid \right)\left(\min_{\psi_2\in \mathcal{M}_1} \sum_{j,l} \mid\la j|u|\psi_2\ra\la \psi_2|{u^\dagger}|l\ra\mid \right)\cr 
&=& 2^2 - (2-{D^1}_k(u))^2.
\ea
 Generalization of this calculation now leads to the proof of the theorem. \\

{\bf Corollary:}  With a similar argument, one can directly derive the following relation:

\be
{D^1}_{{K}}\left(\bigotimes_{i=1}^n u_i\right)\geq 2^n -  \prod_{i=1}^n\left[2-{D^1}_{{k}}(u_i)\right].
\ee

{\bf Physical implications:} Consider  figure (\ref{fig:fig1}) which shows the decohering power of a unitary $u$ in various bases. It is seen that except in very rare situations, the gate $u$ is decohering (has a non-zero decohering power).  In view of the above corollary, we will see that the decohering power of the tensor product of such gates, when the number of gates increases, tends to the maximum value, no matter what types of gates and what type of basis we consdier. In fact we have that almost always (i.e. for almost all $u_i$'s and all bases) 

\be
\lim_{n\lo \infty}\frac{{D^1}_{{K}}\left(\bigotimes_{i=1}^n u_i\right)}{2^n-1}=1.
\ee

This means that no matter how individual qubits interact with their environment (even in a unitary way), the interaction of a large number of qubits  with the environment (even in a unitary way), almost always leads to the de-coherence of the state of the qubits, when the coherence of the state is measured in almsot any basis.  \\

 \subsection{CNOT gate}
We now turn to the CNOT gate which is not a tensor product of two gates. We remind that $CNOT|i,j\ra=|i,i+j\ra$ in the computational basis. Since, here we are dealing with pure states, we use the measure () for the $C_K$ coherence of a state $|\psi\ra$ which in this case reduces to the covariance of the observable $K$ in the state $|\psi\ra$. Let us determine the cohering power of the CNOT gate in a few bases. If $K=\sigma_z\otimes \sigma_z$, 
it is obvious that CNOT has no cohering power. It simply changes an incoherent basis state $|0,0\ra$ to an incoherent state $|0,0\ra$ and so on. Similarly if $K=\sigma_x\otimes \sigma_z$, then again the CNOT gate acts as $\{|+,+\ra, |+,-\ra, |-,+\ra, |-,-\ra\}\ \ \lo\ \ \{|+,+\ra, |-,-\ra, |-,+\ra, |+,-\ra\} $, again producing no coherence. It appears that the highest coherence is produced in mixed bases like $\sigma_x\otimes \sigma_z$ or $\sigma_x\otimes \sigma_y$. Consider the former basis. When acting on an incoherent state like $|+,0\ra$, CNOT will produce 
\be
CNOT|+,0\ra=\frac{1}{\sqrt{2}}(|00\ra+|11\ra)=\frac{1}{2}\left(|+,0\ra+|-,0\ra+|+,1\ra-|-,1\ra\right),
\ee   
the last equality showing that the resulting state has the largest variance and hence the largest coherence in the $\sigma_x\otimes \sigma_z$ basis. This is true for other basis states of $\sigma_x\otimes \sigma_z$ like $|-,0\ra$ etc.It is intresting that the state thus obtained has also the largest amount of entanglment, a property which is independent of basis. \\

\section{Summary}
In this work we have done the following: 1) introduced measures for cohering and de-cohering powers of any quantum channel, 2) have shown that optimizations required in the above definitions can be greatly simplifed by restricting the states only to pure ones, 3) have used two concrete measures of coherence for states, have calculated the cohering and de-cohering power of a number of 1, 2 and n-qubit channels, and finally 4) have proved simple formulas which relate the cohering and decohering power of tensor products of unitary gates to those of individual gates. \\

Interestingly we have found that quantum channels, can have unexpected behavior in creating or destroying coherence. These results and insights have been made possible by looking at the concept of coherence from a  quantitative  angle, measured by concise formulas. It will be interesting to extend these results to more complex channels, specially to two-qubit channels. Quite recently new results have appeared which relate a measures of basis-independent coherence to measures of discord and entanglement for two qubit states and put them in a hierarchal structure \cite{yao}. It will then be interesting to investigate such a hierarchy for two qubit channels in their power for creating coherence, discord and entanglement.    

\section*{Acknowledgment}
A. Mani would like to thank National Elite Foundation of Iran for the financial support.
\hspace{.3in}


\appendix

\section*{Appendix: Derivation of equation (\ref{bitDecohFinal}) }\label{ApenA}

\setcounter{equation}{0}
\renewcommand{\theequation}{A\arabic{equation}}

In this appendix we find the minimum value of the function 
\be
F_{\a,\b}(\xi)=\left(1- \sqrt{\alpha (1-\xi)} \right) \left(1- \frac{\beta \xi}{1-\alpha+ \alpha \xi }\right),
\ee
where  $\alpha= 4 p(1-p)$, $\beta= 4p^2(\k \cdot \x)^2$ and $\xi := (\m \cdot \x)^2$, with $\m$ being all unit vectors that are perpendicular to $\k$. This calculation will give the de-cohering power of the bit-flip channel. 
Note that the minimum value of $\xi$ is zero while its maximum acceptable value depends on $\k=(\cos\theta,\ \sin\theta\ \sin\phi,\  \sin\theta\ \cos\phi)^\intercal$. In order to find this maximum value we use the following parametrization for the unit vector $\m$ which should satisfy $\m\cdot\k=0$:
\be
\m=\left(\begin{array}{ccc}
\sin\theta \sin\Omega \cr
-\cos\theta\ \sin\phi\ \sin\Omega - \cos\phi\ \cos\Omega \cr
-\cos\theta\ \cos\phi \sin\Omega+ \sin\phi\ \cos\Omega
\end{array}
\right).
\ee
\\
Using the above representations for $\k$ and $\m$, we easily see that:
\ba
\xi_{\max}&:=&\max (\m \cdot \x)^2 \cr
&=&{1-\cos^2\theta}\cr
&=& 1- (\k \cdot \x)^2,
\ea
and hence we should perform the minimization (\ref{bitDecohPow1}) in the interval $\xi \in [0, \xi_{\max}]$. \\

Differentiating from $F_{\alpha, \beta}(\xi)$ with respect to $\xi$ shows that the optimal points of this function are as follows:
\be
\xi_{1}=-\frac{1-\alpha}{\alpha}, \ \ \ \ \  \ 
\xi_{2,3}=\frac{ -(1-\alpha)(\alpha-\beta)(\alpha-2\beta) \pm \sqrt{-(1-\alpha)(\alpha-\beta)^3 \beta}}{\alpha (\alpha-\beta)^2}.
\ee
Therefore, with regard to the definitions of the parameters $\alpha$ and $\beta$, the necessary and sufficient condition for the function $F_{\alpha,\beta}(\xi)$ to have a positive optimal point is $\beta \geq \frac{1}{2}(\alpha +\sqrt{\alpha})$. That is equivalent to the following constraint on the value of $(\k \cdot \x)^2$:
\be \label{constraint}
(\k \cdot \x)^2 \geq A, \ \ \ \ \ \ \ \ \ A=\frac{1}{2} \left( \frac{1-p}{p} + \frac{\sqrt{4p(1-p)}}{4p^2} \right).
\ee
Hence, if $(\k \cdot \x)^2 < A$ the function $F_{\alpha,\beta}(\xi)$ has no optimal point in the interval $[0, \xi_{\max}]$ and it can easily be seen that its minimum value is achieved in $\xi=0$:
\be\label{E1}
\min_{\xi} F_{\alpha,\beta}(\xi)=1-\sqrt{\alpha}, \ \ \ \ \ \ \ \ \ \ \  \text{if\ \ } (\k \cdot \x)^2 < A
\ee
On the other hand a little algebra shows that when the condition $(\k \cdot \x)^2 \geq A$ is satisfied, $\xi_3$ does not lie in the interval $[0, \xi_{\max}]$ and the minimum value of $F_{\alpha,\beta}(\xi)$ will be attained in the point  $\xi_{\max}$:
\be\label{E2}
\min_{\xi} F_{\alpha,\beta}(\xi)=\frac{1-4p (\k\cdot\x)^2 (1-p (\k\cdot\x)^2)}{1+\sqrt{4p(1-p)(\k\cdot\x)^2}} , \ \ \ \ \ \ \ \ \ \ \  \text{if\ \ } (\k \cdot \x)^2\geq A
\ee
Gathering the results of equations (\ref{bitDecohPow1}), (\ref{E1}) and (\ref{E2}), we find the $\k-$decohering power of the bit-flip channel as in equation (\ref{bitDecohFinal}).

\end{document}